\documentclass{amsart}

\usepackage{amssymb}

\let\ge\geqslant
\let\le\leqslant

\begin{document}

\pagestyle{plain}

\begin{abstract}
A model of quantum computing is presented, based on
properties of connections with a prescribed monodromy group on
holomorphic vector bundles over bases with nontrivial topology.
Such connections with required properties appear in the
WZW-models, in which moreover the corresponding $n$-point
correlation functions are sections of appropriate bundles which
are holomorphic with respect to the connection.
\end{abstract}

\title{Universal quantum computing based on monodromy representations}

\author{G. Giorgadze}

\maketitle

\subsection*{Logical gates for a quantum computer} Quantum mechanical processes give rise
to new types of computation. Computational networks built out of quantum-mechanical gates
provide a natural framework for constructing quantum computers. The computing capacity of
such computers drastically exceeds that of traditional computer \cite{deutsch}, \cite{shor1}.

The quantum analogue of the classical bit is the quantum bit or
\emph{qubit}. Just as the classical bit is represented by a system
which can adopt one of two distinct states `$0$' and `$1$', one
can define a quantum bit as follows:

\

\noindent{\bf Definition 1.} A \emph{qubit} is a quantum system
whose state can be fully described by a superposition of two
orthogonal eigenstates labeled $|0\rangle$ and $|1\rangle$.

\

The space of quantum states is a Hilbert space, which we denote by $\mathfrak H$. Therefore, a qubit is a normalized state in the two-dimensional Hilbert space ${\mathbf C}^2$. A general state $|\Psi\rangle\in{\mathfrak H}$ of the qubit is given by $|\Psi\rangle=\alpha|0\rangle+\beta|1\rangle$ with $|\alpha^2|+|\beta^2|=1$.

The value of the qubit is the observable $N$ with the Hermitian operator $N|i\rangle=i|i\rangle$ on the Hilbert space ${\mathfrak H}\cong{\mathbf C}^2$, or, in the matrix representation
$$N=
\begin{pmatrix}
0&0\\
0&1\\
\end{pmatrix}.
$$
The expectation value of $N$ is given by
$$
\langle N\rangle=\langle\Psi|N|\Psi\rangle=(\alpha^*\beta^*)
\begin{pmatrix}
0&0\\
0&1\\
\end{pmatrix}
\begin{pmatrix}
\alpha\\
\beta\\
\end{pmatrix}=|\beta|^2
$$
Thus, $\langle N\rangle$ gives the probability to find the system in state $|1\rangle$ if a measurement is performed on the qubit.

While the state of a classical computer can be given as the collection of distinct states of all bits in the memory and processor registers, the ``state of a qubit'' is a meaningless term, if the machine state is the combined state of more than one system.

\

\noindent{\bf Definition 2.} The machine state $|\Psi\rangle$ of
an $n$-qubit quantum computer is given by
$|\Psi\rangle=\sum_{(d_0,...,d_{n-1})}c_{d_0,...,d_{n-1}}|d_0...d_{n-1}\rangle
$ with $\sum|c_{d_0,....d_{n-1}}|^2=1$.

\

The quantum state space ${\mathfrak H}$ is thus the tensor product of $n$ single qubit Hilbert  spaces ${\mathfrak H}_j\cong{\mathbf C}^2$, i.~e.
\begin{equation}\label{tensor}
{\mathfrak H}\cong{\mathfrak H}_1\otimes{\mathfrak H}_2\otimes...\otimes{\mathfrak H}_n\cong({\mathbf C}^2)^{\otimes(n)}
\end{equation}

One of the fundamental facts of the classical theory of computing
is that there exist finite sets of simple functions, called
\emph{fundamental} or \emph{universal gates}, such that any
function $f:{\mathbf B}^n\to{\mathbf B}^m$, where ${\mathbf B}$ is
the Boolean algebra $\{0,1\}$, can be constructed in a simple
explicit way from them. For example the set \{NOT, OR, AND\} is
such a basis of classical computation.

Similarly, one considers \emph{quantum gates} on $k$ qubits (or
$k$-gates) --- unitary $2^k\times 2^k$-matrices acting on the
quantum state space of $k$ qubits. A fundamental problem in
quantum computing is to find basis of gates, which is
``universal''. (More precisely, one distinguishes between
universal bases and more restrictive \emph{exactly universal}
ones; a basis $B$ is called exactly universal if, for each
$k\ge2$, every unitary $k$-qubit operator can be obtained exactly
by a circuit made up of the $k$-qubit gates produced from the
elements of the basis $B$. See \cite{brylinski})

In real world computation, important r\^ole play devices
independent from the environment noise. Thus two of the main
requirements on error-free operations are to have a set of gates
that is both universal for quantum computing and that can operate
in noise-producing environment, i.~e. is fault-tolerant (see
\cite{boykin}). For gates involving irrational multiples of $\pi$,
called non-elementary, fault-tolerant realization is impossible.
Thus, presence of this property makes an elementary gate
inappropriate for physical realization.

It is feasible that any quantum system which one would consider
potentially useful for quantum computing should contain a
generating set of gates, which forms a basis in the above sense.

Two bases $A$ and $B$ are called equivalent, if the gates in the
basis $A$ can be exactly realized using gates in the basis $B$ and
vice versa.

Let us introduce the following operators:
$$
\sigma_x=
\begin{pmatrix}
    0&1\\
    1&0\\
    \end{pmatrix},
\sigma_z=
\begin{pmatrix}
    1&0\\
    0&-1\\
    \end{pmatrix},
\sigma_y=
\begin{pmatrix}
    0&-i\\
    i&0\\
    \end{pmatrix},
\sigma_z^{\alpha}=
\begin{pmatrix}
    1&0\\
    0&e^{i\pi\alpha}\\
    \end{pmatrix},
$$
where $\sigma_x,\sigma_y,\sigma_z$ are Pauli matrices and have the property: every traceless and Hermitian $2\times 2$ unitary matrix $U$ can be expressed as $U=x\sigma_x+y\sigma_y+z\sigma_z,$ where $x,y,z\in{\mathbf R}$ and $x^2+y^2+z^2=1$.

For every unitary operator $U\in{\mathrm U}(2)$ we define the controlled $U$-operator as
$$
\Lambda_k (U)|x_1,...,x_k\rangle \otimes |\xi \rangle
=\begin{cases}
    |x_1,...,x_k\rangle\otimes U|\xi\rangle &\textrm{if }\bigwedge_{j=1}^kx_j=1\\
    |x_1,...,x_k\rangle\otimes  |\xi\rangle &\textrm{if }\bigwedge_{j=1}^kx_j=0
\end{cases}
$$
for all $x_1,...,x_k \in{\mathbf B}$. Here $\bigwedge_{j=1}^kx_j$
denotes action of the operator AND on the boolean variables
$\{0,1\}$. For example, if $k=0$, then
$\Lambda_0(\sigma_x)=\mathrm{NOT}_q:H\to H$, which is called the
\emph{quantum NOT} operator and by definition one has
$\sigma_x=\mathrm{NOT}_q$, which acts on the state
$\alpha|0\rangle+\beta|1\rangle\in{\mathfrak H}$ via
$\alpha|0\rangle+\beta|1\rangle\mapsto\beta|0\rangle+\alpha|1\rangle$.
If the quantum state $\alpha|0\rangle+\beta|1\rangle$ is written
in vector form as $(\alpha,\beta)^\top$, then
NOT$_q(\alpha,\beta)=(\beta,\alpha)$. Second important example is
the so called \emph{controlled-not} operator --- cNOT: ${\mathfrak
H}\otimes{\mathfrak H}\to{\mathfrak H}\otimes{\mathfrak H}$, which
acts on the pairs of qubits and carries out bitwise summation
cNOT$|u,v\rangle=|u,v\oplus u\rangle$, where $\oplus$ is addition
mod 2. Let us consider one more operator --- the \emph{controlled
controlled NOT} operator ccNOT: ${\mathfrak H}\otimes {\mathfrak
H}\otimes{\mathfrak H}\to{\mathfrak H}\otimes{\mathfrak
H}\otimes{\mathfrak H}$.  The classical one and two bit operators
OR and AND can be expressed via ccNOT as follows:
$$
\begin{aligned}
\mathrm{ccNOT}|1,1,a\rangle&=|1,1,(\mathrm{ NOT\ } a)\rangle,\\
\mathrm{ccNOT}|a,b,0\rangle&=|a,b,(a \mathrm{\ AND\ } b)\rangle,
\end{aligned}
$$
for $a,b\in{\mathbf B}$.

On the other hand, every Boolean function can be expressed in the basis \{NOT, OR, AND\} and therefore there exists the unitary operator
$$
U_f:{\mathfrak H}\otimes...\otimes{\mathfrak H}\to{\mathfrak H}\otimes...\otimes{\mathfrak H}
$$
such that $U_f|x,0\rangle=|x,f(x)\rangle$.

Two and three qubit operations, respectively, cNOT and ccNOT, are represented by one qubit operations as follows:
$$
\begin{aligned}
\mathrm{cNOT}&=|0\rangle\langle1|\otimes\mathbf1+|1\rangle\langle1|\otimes\mathrm{NOT},
\mathrm{ccNOT}&=|0\rangle\langle0|\otimes\mathbf1\otimes\mathbf1+|1\rangle\langle1|\otimes
\mathrm{cNOT}.
\end{aligned}
$$
This means that if it is possible to find a finite subset of U(2)
which generates a dense subset of U(2), then we obtain all
operators for quantum computing.

The universal sets of gates for computation have been extensively
studied. To emphasize importance of such works let us recall
some fundamental facts which concern relations between different
bases.

The set of gates $\{\Lambda_2(\sigma_x),\sigma_z^{1/2},H\}$, where
$$
H=\frac1{\sqrt2}\begin{pmatrix}
    1&1\\
    -1&1\\
\end{pmatrix},
$$
the so called Hadamard operator, is a fault-tolerant basis
\cite{shor}. On the other hand the set of gates
$\{\Lambda_1(\sigma_z^{1/2}),H\}$ is universal \cite{Kitaev} and
this basis is equivalent to the basis $\{\Lambda_2(\sigma_x),
\sigma_z^{1/2}, H\}$. In \cite{boykin} is considered the basis
$\{\Lambda_1(\sigma_x), \sigma_z^{1/4}, H\}$ and proved that it is
not equivalent to the basis $\{\Lambda_2(\sigma_x),
\sigma_z^{1/2}, H\}$.

\subsection*{Connection with singularity as generator of quantum gates}
In \cite{zanardi} is considered the model of quantum computing,
called by authors holonomic, based on the well known Berry phase.
Main ingredient in holonomic quantum computation is a smooth
vector bundle $E\to X$ with fibre ${\mathbf C}^2\otimes...\otimes
{\mathbf C}^2$ and unitary connection $\Omega$. The encoding space
of information is in this case the fibre of the bundle and
processing of information is represented by the holonomy operator
$P\exp(\int_\gamma\Omega)$, which acts on the encoding space as
$v\mapsto P\exp(\int_\gamma\Omega)v$, where $\gamma:[0,1]\to X$ is
a smooth path and $P$ denotes path-ordered exponential.

In the present section is given the construction of a universal
set of gates from the monodromy representation of any system of
differential equations of Fuchs type, which can be considered as
dynamic equation of a quantum system. By our opinion, sets of
gates obtained in this way are expected to be fault-tolerant by
reasons similar to ones given in \cite{Freedman}: action of a
monodromy operator corresponding to a loop is unchanged by small
fluctuations of the loop, provided they do not result in crossing
any singularities.

At first let us review a simple example, the differential equation
with regular singular points (for more details see
\cite{bolibruch}) $\frac{df}{dz}=\frac{a}{z}f$, $a\in {\mathbf
C}$, on ${\mathbf C}-\{0\}$. The solution of this equation is the
many valued function $f(z)=z^a$ which by definition is
$z^a=e^{a\log z}$, which under analytic continuation along a path
$\gamma$ looping once counterclockwise around the origin
transforms into $e^{2\pi i a}z^a$. The fundamental group of
${\mathbf C}-\{0\}$ is isomorphic to ${\mathbf Z}$, with generator
$\gamma$, and the monodromy representation in GL$_1({\mathbf C})$
is given by $\gamma\mapsto e^{2\pi ia}$. Analogously, let $A$ be a
Hermitian $2\times 2$ constant matrix, then the matrix function
$(z-s)^A$ is a many valued function on the domain $|z-s|>0$ and is
the solution of the matrix differential equation
$\frac{dF}{dz}=\frac AzF$. The monodromy representation of this
system is given by $\gamma\mapsto e^{2\pi iA}$, and as $A$ is
Hermitian, $e^{2\pi iA}$ is unitary and therefore we obtain the
monodromy representation in the unitary group. In particular, if
we can obtain any one-qubit gate, for example $\tau$, it will
suffice to choose a Hermitian $2\times 2$ matrix $\alpha$, such
that $e^{2\pi i\alpha}=\tau$, then $e^{2\pi i\alpha}$ acts as
locally unitary operator (processing of information)  on  the
fibre (encoding space of information) ${\mathbf C}^2$ of the
trivial vector bundle ${\mathbf C}^2\times{\mathbf
C^{*}}\to{\mathbf C^{*}}$.

In \cite{giorgadze} we consider the case, when the holomorphic
vector bundle is given on a punctured compact Riemann surface. In
this case, fundamental group of the base of the bundle is a free
group and obtaining a universal set of logical gates does not
present a difficult problem after application of solution methods
of the Riemann-Hilbert problem.

Let $\mathbf{CP}^n$ be the $n$-dimensional complex projective
space, and let $s_1,s_2,s_3,s_4\in\mathbf{CP}^1$ and $s_j\neq \infty$. Denote by
$X_4=\mathbf{CP}^1-\{ s_1,s_2,s_3,s_4 \}$ and by
$\gamma_1,\gamma_2,\gamma_3,\gamma_4$ the generators of
$\pi_1(X_4,z_0)$ with relations
$\gamma_1\gamma_2\gamma_3\gamma_4=1$. Let $M_1,M_2,M_3,M_4$ be
matrices such that $E_j=\frac1{2\pi i}\ln M_j, j=1,2,3,4$ and
$M_1,M_2,M_3$ generate a basis of the Lie algebra of SU(2) and
satisfy the condition  $M_1 M_2 M_3 M_4=1$. Consider the
representation $\rho : \pi_1(X_4,z_0)\rightarrow\mathrm{SU}(2)$ defined by
the map $\gamma_j \mapsto M_j$. Then for $\rho$ there exits a
system of differential equations of Fuchs type
$$
d F(z)= (\sum_{j=1}^4 \frac {A_j}{z-s_j}dz)F(z),
$$
whose monodromy representation coincides with $\rho$ \cite{bolibruch}. The monodromy representation $\rho$ induces two-dimensional vector bundle $E \rightarrow \mathbf{CP}^1$ with meromorphic connection form
 $$
 \omega= \sum_{j=1}^4 \frac {A_j}{z-s_j}dz
 $$
and Chern number $c_1(E)=\sum_{j=1}^4 E_j$. The solution space
$\mathfrak H$ of the system is a two-dimensional vector space.
Moreover $\pi_1(X_4,z_0)$ acts on $\mathfrak H$ and any unitary
operator can be obtained in this way.

Therefore, we have proved the following proposition:

\ 

\ \noindent{\bf Proposition 1} \cite{giorgadze}. {\sl The
connection form $\omega$ generates the basis for the computation.}

\ 

Consider the general case. Let $X$ be a compact Riemann surface of
genus $g \geq 2$ with marked point $z_0$. The fundamental group of
$X ^\prime = X-\{ z_0 \}$ is generated by
$\alpha_1,\beta_1,...,\alpha_g,\beta_g, \gamma$ with relations
$\Pi_{j=1}^{g} [\alpha_j,\beta_j]=\gamma$, where
$\alpha_1,\beta_1,...,\alpha_g,\beta_g$ are generators of the
fundamental group $\pi_1(X)$ and $\gamma$ is a loop going around
$z_0$. Consider such homomorphism $\rho: \pi_1(X \prime)
\rightarrow$ SU(2) that Im$\rho$ is a dense subgroup of SU(2).
The homomorphism $\rho$ defines a two-dimensional holomorphic
vector bundle $E^{\prime}\rightarrow X^{\prime}$ and there exists
a SU(2)-system of differential equations $Df=\omega f$ on $X$
which has one regular singular point $z_0$ and the monodromy
representation of this system coincides with $\rho$. The pair
$(E^{\prime},\omega)$ can be extended to a possibly
holomorphically nontrivial bundle $E\rightarrow X$, for which
$\omega$ can be a meromorphic connection. Therefore we proved the
following proposition

\

\noindent{\bf Proposition 2} \cite{giorgadze}. {\sl The connection
$\omega$ of the bundle $E\rightarrow X$ densely generates all unitary
operators $\mathfrak H \rightarrow \mathfrak H$}.

\ 

Below we  consider the case when the fundamental group of the base is not
free.

Suppose $D=\bigcup_{j=1}^{m+1}D_j$ is a divisor, where
$D_j$, $j=1,...,m+1$ are hyperplanes in $\mathbf{CP}^n$. The main
result of this paper is the following theorem.

\

\noindent{\bf Theorem}. {\sl There exists a Fuchs type Pfaff
system
$$
df=\omega f
$$
on $\mathbf{CP}^n-D$ whose monodromy representation gives a universal set
of quantum gates.}

\

Let us choose a line $L$ which intersects the divisor
$D=\bigcup_{j=1}^{m+1}D_j$ at nonsingular points $a_j=L\cap D_j$,
$j=1,2,...,m+1$. Consider the fundamental group
$\pi_1(L-\{a_1,...,a_{m+1}\},z_0)$. Let
$[\gamma_1],...,[\gamma_m]$ be the generators of
$\pi_1(L-\{a_1,...,a_{m+1}\},z_0)$, where the loops have form
$\gamma_j=\sigma_j[\alpha_j]\sigma_j^{-1}$, $\alpha$ is a path
from $z_0$ to a neighborhood $V_{a_j}$ and $\sigma_j$ is small a
loop in the the neighborhood $V_{a_j}$ which generates
$\pi_1(V_{a_j}-\{a_j\})\cong{\mathbf Z}$.

It is known that $[\gamma_1],...,[\gamma_m]$ too are generators of
$\pi_1(\mathbf{CP}^n-D,z_0)$ under some conditions.

Suppose we have the family of representations
\begin{equation}\label{repfamily}
\rho_\lambda:\pi_1(\mathbf{CP}^n-D,z_0)\to\mathrm{GL}_m({\mathbf C})
\end{equation}
such that
\begin{equation}\label{series}
\rho_\lambda([\gamma_j])=1+\lambda M_1^j+\lambda^2M_2^j+...+\lambda^kM_k^j+...,
\end{equation}
where $M_k^j$ are $m\times m$-matrices. The family of
representations (\ref{repfamily}) satisfying the condition
(\ref{series}) is called analytic.

Suppose $D_j=\{h_j=0\}$ and consider the 1-form
$\Omega_k=\sum_{j=1}^{m+1}U_k^j\frac{dh_j}{h_j}$, where $U_k^j$
are constant matrices and $\sum_{j=1}^{m+1}U_k^j=0$. Consider the
family of meromorphic 1-forms
\begin{equation}\label{omega}
\Omega(\lambda)=\lambda\Omega_1+\lambda^2\Omega_2+...+\lambda^k
\Omega_k+... .
\end{equation}

If $U^j(\lambda)=\sum_{k=1}^{\infty}\lambda^k U_k^j$ is a
converging power series then the family $\Omega(\lambda)$ is
called analytic family of Fuchs systems.

It is known that for every $\lambda$ this system satisfies the condition
$$
d_z\Omega(\lambda)=0, z\in C.
$$
If the analytic family of Fuchs systems satisfies the condition
$$
\Omega(\lambda)\wedge\Omega(\lambda)=0,
$$
then the family of Fuchs systems is called integrable.

Similar terminology will be used for a Pfaff system
$df=\Omega(\lambda)f$.

The monodromy representation of the integrable family of Pfaff systems
\begin{equation}\label{paff}
df=\Omega(\lambda)f
\end{equation}
is an analytic family of representations of the fundamental group
$\pi_1(\mathbf{CP}^n-D,z_0)$ \cite{Leksin}.

Let $\Phi$ be the fundamental matrix of the system (\ref{paff})
represented by the Peano series
\begin{equation}\label{piano}
\Phi=1+\int \Omega(\gamma)+\int
\Omega(\gamma)\Omega(\gamma)+...+\int
\Omega(\gamma)...\Omega(\gamma)+...,
\end{equation}
where $\int \Omega(\gamma)...\Omega(\gamma)$ is the Chen iterated integral. If such representation of the fundamental matrix is chosen, then for every $\gamma \in \pi_1(\textbf{CP}^n-D,z_0)$ one will have
$$
\rho([\gamma])=1+\int_{\gamma} \Omega(\gamma)+\int_{\gamma}
\Omega(\gamma)\Omega(\gamma)+...+\int_{\gamma}
\Omega(\gamma)...\Omega(\gamma)+....
$$

\

\noindent{\bf Theorem 1} \cite{Leksin}. {\sl For every analytic
family of representations (\ref{repfamily}), when the parameter is
small, there exists a family of Pfaff systems (\ref{paff}) whose
monodromy coincides with (\ref{repfamily}).}

\ 

\noindent{\it Sketch of proof.} We will show, that there exists a
family of Fuchs type Pfaff systems (\ref{paff}), where
$\Omega(\lambda)$ has the form (\ref{omega}),
$\Omega_k=\sum_{j=1}^{m}U_k^j\omega_j$,
$\omega_j=\frac{dh_j}{h_j}-\frac{dh_{m+1}}{h_{m+1}}$, so that its
family of representations coincides with (\ref{repfamily}).

Step 1. By (\ref{repfamily}) we find $U_k^j$, $k=1,2,...$,
$j=1,2,...,m$. Indeed, let $\rho_\lambda([\gamma_j])$ be
represented as Peano series like (\ref{piano}), and rewrite it in
Lappo-Danilevski form \cite{golubeva}:
\begin{equation}\label{lappoform}
\rho([\gamma])=1+\sum_{j=1}^{m}\int_\gamma \omega_j U^j(\lambda)+
\sum_{j,k=1}^{m}\int_\gamma \omega_j\omega_k
U^j(\lambda)U^k(\lambda)+...+
\end{equation}
$$
+\sum_{j_1,...,j_k=1}^m\int_\gamma\omega_{j_1}...\omega_{j_k}
U^{j_1}(\lambda) U^{j_k}(\lambda);
$$
for the generators we have (\ref{repfamily}) and therefore we
obtain:
$$
\lambda M_1^j+\lambda^2 M_2^j+...+\lambda^k M_k^j+...=
$$
$$
=\lambda\int_{\gamma_j}\Omega_1+\lambda^2(\int_{\gamma_j}\Omega_1\Omega_1+\int_{\gamma_j}\Omega_2)+...
$$
$$
...+\lambda^k(\int_{\gamma_j}\Omega_k+\sum_{q=2}^{m}\sum_{k_1+...+k_q=k}\int_{\gamma_j}\Omega_{k_1}...
\Omega_{k_q})+...
$$
This implies
$$
\int_{\gamma_j}\Omega_1=M_1^j, j=1,...,m
$$
$$
\int_{\gamma_j}\Omega_1\Omega_1+\int_{\gamma_j}\Omega_2=M_2^j,
j=1,...,m,
$$
$$
\int_{\gamma_j}\Omega_k+\sum_{q=2}^{m}\sum_{k_1+...+k_q=k}\int_{\gamma_j}\Omega_{k_1}...
\Omega_{k_q}=M_k^j, j=1,...,m
$$
and so on. As $\int_{\gamma_j}\Omega_k=2\pi i U_k^j$, for $U_k^j$,
$j=1,...,m, k=1,2,...$ we obtain
$$
U_1^j=\frac1{2\pi i}M_1^j,
$$
$$
U_2^j=\frac1{2\pi
i}(M_2^j-\int_{\gamma_j}\Omega_1\Omega_1)=\frac1{2\pi i
}(M_2^j-\sum_{k_1,k_2=1}^m\int_{\gamma_j}
\omega_{k_1}\omega_{k_2}U_1^{k_1}U_1^{k_2}),
$$
$$
..........
$$
$$
U_k^j=\frac1{2\pi i }(M_k^j-\sum_{q=2}^k
\sum_{k_1+k_2+...+k_q=k}\int_{\gamma_j}
\Omega_{k_1}...\Omega_{k_q}),
$$
$ j=1,2,...,m. $ Therefore we obtain a formal family of 1-forms
\begin{equation}\label{ser}
\Omega(\lambda)=\lambda\Omega_1+\lambda^2\Omega_2+...+\lambda^k\Omega_k+...=\sum_{j=1}^{m}U^{j}(\lambda)\omega_j,
\end{equation}
$$
\Omega_k=\sum_{j=1}^{m}U_{k}^{j}\omega_j,
U^j(\lambda)=\sum_{k=1}^{\infty}\lambda^k U_k^j.
$$

Step 2. The formal series (\ref{ser}) is convergent for small
$\lambda$ \cite{Leksin}.

Step 3. The family of 1-forms (\ref{ser}) is integrable, i.~e. the
identity $\Omega(\lambda)\wedge\Omega(\lambda)=0$ is satisfied
\cite{Leksin}.

\

\noindent{\bf Theorem 2}. {\sl Let
$$
\rho:\pi_1(\mathbf{CP}^n-D,z_0)\to\mathrm{GL}_m({\mathbf C})
$$
be a representation such that $\rho([\gamma_j])$, $j=1,...,m$, are
close to identity. Then $\rho$ is realizable as monodromy
representation of an integrable Fuchs system $df=f \Omega$,
$\Omega=\sum_{j=1}^{m}U^{j}\omega_{j}$, where $U^j$ are close to
the zero matrix}.

We will apply results of this subsection to a special divisor. In
particular, suppose $D=\bigcup_{i<j}D_{ij}\cup D_0$, where
$D_{ij}=\{(z_0,...,z_n)\in\mathbf{CP}^n | z_i=z_j, i,j\neq 0 \}$,
$D_0=\{z\in\mathbf{CP}^n | z_0=0 \}$. Let $X_n=\mathbf{CP}^n-D=\{
(z_1,...,z_n)\in{\mathbf C}^n|z_i\neq z_j, i\neq j \}$. The
fundamental group of $X_n$ is called the pure braid group on $n$
strings, which we will denote by $P_n$. The symmetric group $S_n$
acts on $X_n$ by $g.(z_1,...,z_n)=(z_{g(1)},...,z_{g(n)})$, $g\in
S_n$. The fundamental group of the quotient space $X_n/S_n$ is
called the braid group on $n$ strings and denoted by $B_n$. The
braid group has $n-1$ generators $\sigma_1,...,\sigma_{n-1}$ with
relations
\begin{equation}\label{braid1}
\sigma_i\sigma_{i+1}\sigma_{i}=\sigma_{i+1}\sigma_{i}\sigma_{i+1},
1\le i \le n-2,
\end{equation}
\begin{equation}\label{braid2}
\sigma_i\sigma_j=\sigma_j\sigma_i, |i-j|\ge 2.
\end{equation}
Choose a point $z_0=(1,2,...,n)\in X_n$ and denote by $pr:X_n\to
X_n/S_n$ the natural projection. We have exact sequence of groups:
$$
1\to P_n \to B_n \to S_n \to 1
$$
and generators of the pure braid group $P_n$ are $\tau_{ij}$, $1\le i
< j\le n$, where
\begin{equation}\label{pure}
\tau_{ij}=\sigma_i\sigma_{i+1}...\sigma_{j-1}\sigma_{j}^2\sigma_{j-1}^{-1}...\sigma_i^{-1},
1\le i < j \le n
\end{equation}
satisfying the relations
$$
\tau_{rs}\tau_{ik}=\tau_{ik}\tau_{rs}, \textrm{ for } s<i \textrm{
or } k<s,
$$
$$
\tau_{ks}\tau_{ik}\tau_{ks}^{-1}=\tau_{is}^{-1}\tau_{ik}\tau_{is}
\textrm{ for } i<k<s,
$$
$$
\tau_{rk}\tau_{ik}\tau_{rk}^{-1}=\tau_{ik}^{-1}\tau_{ir}^{-1}\tau_{ik}\tau_{ir}\tau_{ik}
\textrm{ for } i<r<k,
$$
$$
\tau_{rs}\tau_{ik}\tau_{rs}^{-1}=\tau_{is}^{-1}\tau_{is}\tau_{ir}\tau_{ik}\tau_{ir}^{-1}\tau_{is}^{-1}\tau_{ir}\tau_{is}
\textrm { for } i<r<k<s.
$$

Consider the matrix valued 1-form
 \begin {equation}\label{form}
 \Omega=\sum_{1\le i < j \le n}\Omega_{ij}dlog(z_i-z_j)
 \end{equation}
 holomorphic on $X_n$, where $\Omega_{ij}\in M_m(C), 1\le i < j \le
 n$.
Let $E\to X_n$ be a holomorphic rank $m$ trivial vector bundle with connection
$\nabla$ for which $\Omega$ is connection form. Holomorphic sections $f=(f^1,...,f^m)$ of this bundle are solutions of the Fuchs system
 \begin{equation}\label{fuchs}
 df=\Omega f,
\end{equation}
where $\Omega$ has the form (\ref{form}).

\

\noindent{\bf Proposition 3}. {\sl The system (\ref{fuchs}) is
integrable if and only if the following condition is satisfied}
\begin{equation}\label{condition1}
[\Omega_{ij},\Omega_{ik}+\Omega_{jk}]=[\Omega_{ij}+\Omega_{ik},\Omega_{ji}],i<j<k,
\end{equation}
\begin{equation}\label{condition2}
[\Omega_{ij},\Omega_{kl}]=0, \textrm{ for distinct } i,j,k,l.
\end{equation}

\

By theorem 2 we have \cite{kohno1}:

\

\noindent{\bf Theorem 3}. If $\rho:P_n\to\mathrm{GL}_m({\mathbf
C})$ is such representation that $||\rho(\tau_{ij})-1||$ is
sufficiently small for each $1\le i<j\le n$, then there exist
matrices $\Omega_{ij}\in M_m(\mathbf C)$, $1\le i<j\le n$ close to $0$,
which satisfy the conditions (\ref{condition1})-(\ref{condition2})
and monodromy representation of the Fuchs system
$$
df=\sum_{1\le i<j\le n}\Omega_{ij}\frac{d(z_i-z_j)}{z_i-z_j}f
$$
coincides with $\rho$.

\

Let us apply the above results for designing logical gates for
quantum computing. Choose such $n$ and representation
$\rho:P_n\to\mathrm{SU}(m)$ that, the image Im$\rho(P_n)$ is dense
in SU(m) and $||\rho(\tau_{ij})-1||$ is sufficiently small for
each $1\le i<j\le n$. Then the conditions of theorem 3 are
satisfied and therefore there exists the unitary connection
\begin{equation}\label{connnection}
\nabla=\sum_{1\le i<j\le n}\Omega_{ij}\frac{d(z_i-z_j)}{z_i-z_j}
\end{equation}
on the trivial $m$-dimensional vector bundle ${\mathbf C}^m\times
X_n \to X_n$ which gives the action on ${\mathbf C}^m$ as follows:
${\mathbf C}^m\ni v \mapsto \rho(\tau)v$, for every $\tau \in
P_n$.

In this manner we obtain universal quantum logical gates which are
based on holomorphic vector bundles over $X_n$ with integrable
connection (\ref{connnection}).

\subsection*{Conformal field theory} Consider a
1-form $\Omega$ of special type defined in (\ref{form}). Let
$V_1,...,V_m$ be ${\mathfrak sl}_2$-modules. Put
$V=V_1\otimes...\otimes V_m$.  Let $\{ I_j \}$ be an orthogonal
basis of ${\mathfrak{sl}}_2$ and ${\mathbf c}=\sum_jI_jI_j \in
U(\mathfrak{sl}_2)$ the Casimir element in the universal
enveloping algebra $U(\mathfrak{sl}_2)$. Let $\Delta
:U(\mathfrak{sl}_2) \to U(\mathfrak{sl}_2) \otimes
U(\mathfrak{sl}_2)$ be the diagonal homomorphism determined by
$\Delta(x)=x\otimes1 + 1\otimes x$, $x\in\mathfrak{sl}_2$. Set
$\Omega=\frac12(\Delta\mathbf{c}-\mathbf{c} \otimes 1 -1 \otimes
\mathbf{c})$. Consider a family
$\rho_i:\mathfrak{sl}_2\to\mathrm{End}(V_i)$, $i=1,...,n$ of
irreducible representations and define the representations
\begin{equation}\label{rep2}
\rho_i\otimes\rho_j:\mathfrak{sl}_2\to\mathrm{End}
(V_1\otimes...\otimes V_n)
\end{equation}
by formul\ae
$$
(\rho_i\otimes\rho_j)(x)=I_1\otimes...\otimes\rho_i(x)\otimes...\otimes
I_j\otimes...\otimes I_n+I_1\otimes...\otimes I_i\otimes...\otimes
\rho_j(x)\otimes...\otimes I_n,
$$
where $I_k$ denotes the identity operator acting on $V_k$.

The representations (\ref{rep2}) extend to the universal
enveloping algebra $U(\mathfrak{sl}_2)$, these representations we
denote again by $\rho_i\otimes\rho_j$; thus we have the
representations $\rho_i\otimes\rho_j:U(\mathfrak{sl}_2)\to\mathrm{
End} (V_1\otimes...\otimes V_n)$. Let $\Omega_{ij}=(\rho_i \otimes
\rho_j)\Omega$. The linear operators $\Omega_{ij}:V\to V$, $i<j$,
act as $\Omega\otimes1+1\otimes\Omega$ on $V_i\otimes V_j$ and
trivially on all of the other factors. The Fuchs type Pfaff system
\begin{equation}\label{kz}
\frac{\partial \Psi (z_1,...,z_n)}{\partial
z_i}=\frac1{\lambda}\sum_{j=1,i\neq
j}^{n}\frac{\Omega_{ij}}{z_i-z_j}\Psi, \ \ \ i=1,...,n,
\end{equation}
where $\Psi(z_1,...,z_n)$ is a $V$-valued function on $X_n$, is
called the Knizh\-nik-Za\-mo\-lod\-chi\-kov equation. Here
$\lambda$ is a complex parameter. Solutions of (\ref{kz}) are
covariant constant sections of the trivial bundle $X_n\times V \to
X_n$ with flat connection
$$\sum_{\substack{j=1\\ i\neq j}}^n\frac{\Omega_{ij}}{z_i-z_j}d(z_i-z_j).
$$
It follows from previous subsections that (\ref{kz}) is integrable
if and only if the conditions
(\ref{condition1})-(\ref{condition2}) are satisfied. The monodromy
representation of (\ref{kz}) can be extended to a representation
of the braid group \cite{kohno1}. Let $V_1=V_2=...=V_n\cong
{\mathrm C}^2$, then we obtain that the braid group $B_n$ acts on
$V^{\otimes n}\cong ({\mathbf C}^2)^{\otimes n}$. For quantum
computing an appropriate choice of this action is necessary (see
\cite{Freedman}).

The Knizhnik-Zamolodchikov equation was invented in conformal
field theory. Its solutions describe $n+1$-point correlation
functions on the Riemann sphere in the Wess-Zumino-Witten model.
This model is uniquely determined by choice of a simple Lie
algebra $\mathfrak g$ for every $k>0$, $k\in{\mathbf Z}$. Any
model of the conformal field theory has a certain set of primary
fields $\{\phi_j\}$. They are in one-to-one correspondence with
irreducible representations $\{ \rho_j\}$ of the Kac-Moody algebra
$\widehat{\mathfrak g}$. Let ${\mathfrak g}=\mathfrak{sl}_2({\mathbf C})$. For each nonnegative half integer $j$ there exists a unique irreducible $\mathfrak g$-module
$V_j$, called that of spin $j$, with highest weight $j\alpha$ and
dim$V_j=2j+1$.

\

\noindent{\bf Theorem 4} \cite{knizhnik}, \cite{kohno2}. {\sl The
$n$-point function $\Psi=\langle
\phi_1(z_1)...\phi_n(z_n)\rangle$, where $\phi_j$ are primary
fields, satisfies the system of differential equations (\ref{kz})
and the monodromy representation of this system is unitarizable.}

For example consider the case $n=2$ with

$$
\Omega=\Omega_{12}=\Omega_{21}=\begin{pmatrix}
    1/2&0&0&0\\
    0&-1/2&1&0\\
    0&1&-1/2&0\\
    0&0&0&1/2
\end{pmatrix},
$$
then (\ref{kz}) have the form
\begin{align}\label{par}
\frac {\partial F}{\partial z_1}=\frac1{\lambda} \frac {\Omega_{12}}{z_1-z_2}F,\\
\frac {\partial F}{\partial z_2}=\frac1{\lambda} \frac
{\Omega_{21}}{z_2-z_1}F,
\end{align}
where $F$ is a function on $X_2=\textbf{C}^2-\{ z_1=z_2 \}$ with
values in $V^{\otimes 2 }$, $dim_{\textbf{C}} V =2$. The solutions
to the above system are given by $F(z)=e^{\frac1{\lambda}\ln(z_1-z_2)\Omega}C$, where $C$ is a constant 4-vector. The image of the generator $\sigma_1\in B_2$ by
the monodromy representation of the system (\ref{par})-(20) is $e^{\frac{-\pi i}{\lambda}\Omega}$ \cite{lawrence}, which can be considered as a nontrivial 2-qubit gate.

\

In \cite{kanie} is given an example of such 4-point function, for
which the corresponding monodromy representation  is $\sigma_x$.
In \cite{sierra} is considered the exact solution and
integrability of the reduced BCS model of superconductivity
considered from the conformal field theory point of view.

 \

{\small Georgian Academy of Sciences}

{\small Institute of Cybernetics}

{\small e-mail: giorgadze@rmi.acnet.ge}

\bibliographystyle{plain}

\end{document}